\documentclass[fleqn,onecolumn,usenatbib]{mnras}

\usepackage{newtxtext,newtxmath}

\usepackage[T1]{fontenc}

\DeclareRobustCommand{\VAN}[3]{#2}
\let\VANthebibliography\thebibliography
\def\thebibliography{\DeclareRobustCommand{\VAN}[3]{##3}\VANthebibliography}

\usepackage{graphicx}	

\usepackage{pdflscape}

\newcommand{\fract}[2]{\leavevmode\kern.1em
          \raise.5ex\hbox{\the\scriptfont0 #1}\kern-.1em
    \raise.15ex\hbox{\the\scriptfont0 /}\kern-.08em\lower.25ex\hbox{\the\scriptfont0 #2}}

\newcommand{\half}{{\textstyle\frac{1}{2}}}

\newcommand{\quart}{{\textstyle\frac{1}{4}}}

\newcommand{\colsp}{\mathop{\rm colsp}\nolimits}

\newcommand{\bk}{\mathbfit{k}}

\newcommand{\boldu}{{\mathbfit{u}}}
\newcommand{\boldv}{{\mathbfit{v}}}

\newcommand{\A}{{\mathbfss{A}}}
\newcommand{\B}{{\mathbfit{B}}}
\newcommand{\matB}{{\mathbfss{B}}}
\newcommand{\C}{{\mathbfss{C}}}
\newcommand{\matD}{{\mathbfss{D}}}

\newcommand{\G}{{\mathbfss{G}}}
\newcommand{\I}{{\mathbfss{I}}}

\renewcommand{\L}{{\mathbfss{L}}}

\newcommand{\M}{{\mathbfss{M}}}
\renewcommand{\P}{{\mathbfss{P}}}
\newcommand{\Q}{\mathbfss{Q}}
\newcommand{\U}{\mathbfit{U}}

\newcommand{\Z}{\mathbfss{Z}}

\newcommand{\vdot}{{\boldsymbol{\cdot}}}
\newcommand{\vcross}{{\boldsymbol{\times}}}
\newcommand{\grad}{\mbox{\boldmath$\nabla$}}
\newcommand{\diag}{\mathop{\rm diag}}
\newcommand{\thth}{\hspace{1.5pt}}

\newcommand\Div{\grad\vdot\thth}

\newcommand{\kpar}{k_{\scriptscriptstyle\parallel}}

\newcommand{\bv}{Brunt-V\"ais\"al\"a}

\newcommand{\ri}{{i}}

\renewcommand{\leq}{\leqslant}  
  \renewcommand{\ge}{\geqslant}

\title[Fragility of Alfv\'en waves]{On the Fragility of Alfv\'en waves in a Stratified Atmosphere }

\author[P. S. Cally]{
Paul S. Cally,$^{1}$\thanks{E-mail: paul.cally@monash.edu}
\\
$^{1}$School of Mathematics, Monash University, Clayton 3800, Victoria, Australia
}

\date{Accepted XXX. Received YYY; in original form ZZZ}

\pubyear{2021}

\begin{document}
\label{firstpage}
\pagerange{\pageref{firstpage}--\pageref{lastpage}}
\maketitle

\begin{abstract}
Complete asymptotic expansions are developed for slow, Alfv\'en and fast magneto\-hydro\-dynamic waves at the base of an isothermal three-dimensional (3D) plane stratified atmosphere. Together with existing convergent Frobenius series solutions about $z=\infty$, matchings are numerically calculated that illuminate the fates of slow and Alfv\'en waves injected from below. An Alfv\'en wave in a two-dimensional model is 2.5D in the sense that the wave propagates in the plane of the magnetic field but its polarization is normal to it in an ignorable horizontal direction, and the wave remains an Alfv\'en wave throughout. The rotation of the plane of wave propagation away from the vertical plane of the magnetic field pushes the plasma displacement vector away from horizontal, thereby coupling it to stratification. It is shown that potent slow-Alfv\'en coupling occurs in such 3D models. It is found that about 50\% of direction-averaged Alfv\'en wave flux generated in the low atmosphere at frequencies comparable to or greater than the acoustic cutoff can reach the top as Alfv\'en flux for small magnetic field inclinations $\theta$, and this increases to 80\% or more with increasing $\theta$. On the other hand, direction-averaged slow waves can be 40\% effective in converting to Alfv\'en waves at small inclination, but this reduces sharply with increasing $\theta$ and wave frequency. Together with previously explored fast-slow and fast-Alfv\'en couplings, this provides valuable insights into which injected transverse waves can reach the upper atmosphere as Alfv\'en waves, with implications for solar and stellar coronal heating and solar/stellar wind acceleration. 
\end{abstract}




\section{Introduction}
Magneto\-atmospheric-gravity (MAG) waves result from a combination of three restoring forces: compression, which on its own produces sound waves; buoyancy, that gives rise to gravity waves; and magnetic field, that can produce Alfv\'en waves. Prior to the mid-1980s, exact solutions in the linear case were available for any two of these acting together, but not all three.

It therefore came as a revelation when \cite{ZhuDzh84aa} derived exact solutions in terms of Meijer-G functions for 2D waves in a gravitationally stratified isothermal magneto\-atmosphere with uniform inclined magnetic field. Later, it was pointed out that these solutions may be written in terms of more elementary ${}_2F_3$ hypergeometric functions \citep{Cal01aa,Cal09aa}, but this only adds convenience, not new results.

\cite{ZhuDzh84aa} also set out the theory for the 3D case, for which only series solutions were available. Later, \cite{CalGoo08aa} numerically solved these equations in 3D, with a base acoustic source, discovering a rich set of fast-slow and fast-Alfv\'en mode conversion processes of relevance to the propagation of waves from the solar photosphere to the outer atmosphere. The implementation of the bottom boundary conditions by \citeauthor{CalGoo08aa} was inexact but adequate for the purpose.

Although an infinite isothermal atmosphere with uniform magnetic field is a very crude model of a stellar atmosphere, it exhibits the most important characteristic of density stratification over many scale heights that characterises say the solar photosphere/chromosphere. With the sound speed $c$ being a constant or slowly varying function of height and the Alfv\'en speed $a$ increasing rapidly, magneto\-hydro\-dynamic (MHD) waves generated by photospheric granulation must propagate upward from layers where $a\ll c$ to the upper chromosphere where $a\gg c$ if they are to play a role in outer atmospheric heating. As revealed by the two-dimensional (2D) solutions of \cite{ZhuDzh84aa}, fast and slow MHD waves couple and exchange energy in intermediate layers. In three dimensions (3D) the Alfv\'en wave will also be coupled into the process, affecting the nature of waves that arrive in the upper atmosphere. The isothermal model is the simplest that captures these processes in a mathematically tractable way, and so provides a valuable pedagogical testbed. The steep transition region atop the chromosphere is not addressed in the model, and will produce further reflection and transmission \citep{HanCal12aa}, but nevertheless it is essential to understand which waves reach it and in which form. The fate of Alfv\'en and slow waves excited in the photosphere is our particular topic of interest here.

Here we derive asymptotically exact boundary conditions and apply them to the injection of a pure Alfv\'en or slow wave at the bottom to determine what fraction of wave-energy flux reaches the top, and how much is reflected or mode-converted to magneto-acoustic waves. This gives important insights into the fragility of Alfv\'en waves in stratified atmospheres.

\begin{figure}
\begin{center}
\includegraphics[width=0.4\textwidth]{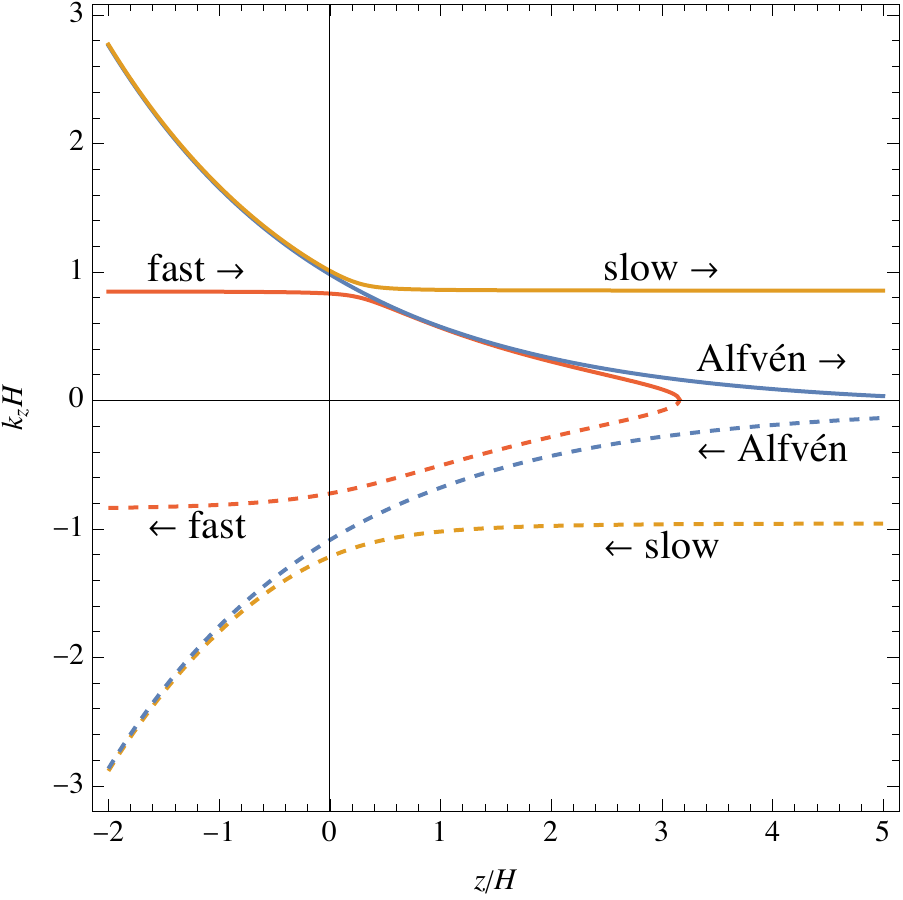}\qquad\qquad\includegraphics[width=0.4\textwidth]{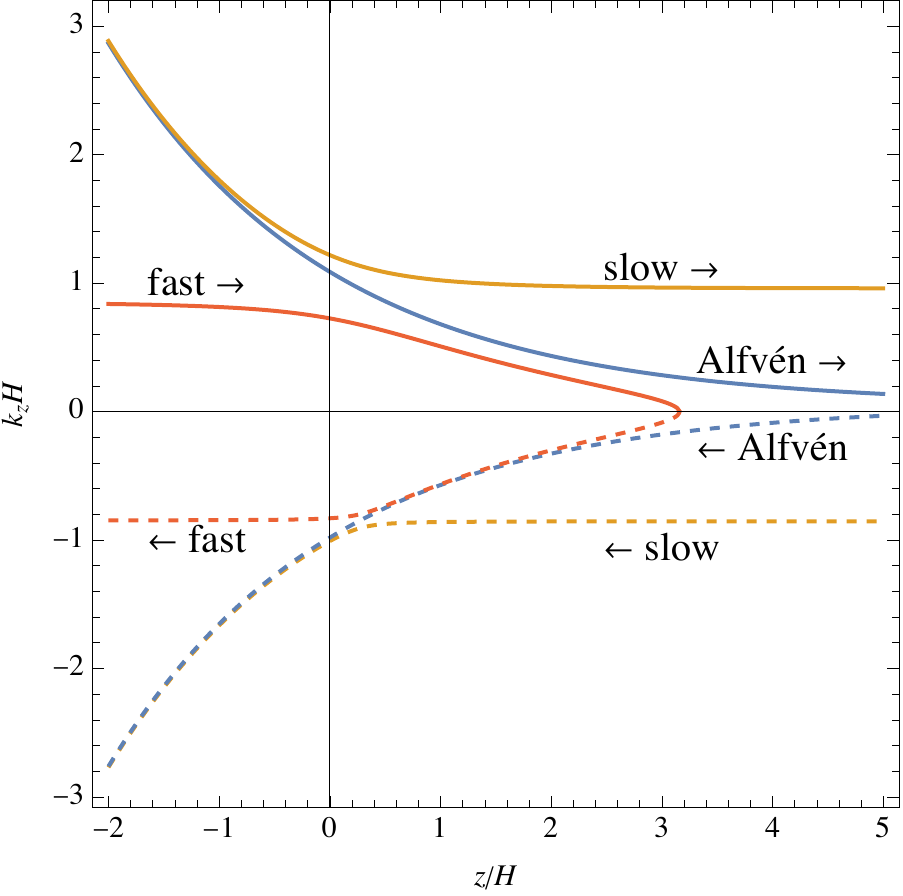}
\caption{Dispersion diagrams in $z$--$k_z$ space for a wave with dimensionless frequency $\nu=\omega H/c=1$ and dimensionless horizontal wavenumber $\kappa=k_xH=0.2$ in an isothermal atmosphere with sound speed $c$ and density scale height $H$, and with magnetic field inclination from the vertical $\theta=15^\circ$ and orientation out of the $x$--$z$ plane $\phi=15^\circ$ (left frame) and $\phi=165^\circ$ (right frame). The Alfv\'en-acoustic equipartition $a=c$ is placed at $z=0$. To be consistent with later figures, the slow wave loci are rendered in orange, Alfv\'en in blue, and fast in red, with dashed curves indicating downward propagation. Based on the dispersion relation of \citet{NewCal10aa}.}
\label{fig:disp}
\end{center}
\end{figure}

Opportunities for coupling are apparent in the typical dispersion diagrams Figure \ref{fig:disp}. The $z$--$k_z$ plane loci of the dispersion curves 
for injected slow or Alfv\'en waves from the bottom are almost coincident until near the Alfv\'en-acoustic equipartition height $z=0$, and so may potentially interact, if there is an available coupling mechanism. We shall show that gravitational stratification combined with magnetic field inclination $\theta$ and orientation $\phi$ away from the vertical plane of propagation provides such a mechanism. This is because the plasma displacement vector $\bxi$ has a component in the $z$-direction, and hence the plasma perturbation interacts with the gravitational stratification. The Hall effect in a weakly ionized plasma enables slow-Alfv\'en coupling too \citep{RabCal19aa,RabCal21lk}, but will not be considered here.

Figure \ref{fig:disp} also illustrates close encounters between the upgoing fast and slow loci near the Alfv\'en-acoustic equipartition level $z=0$, and between the upgoing fast and Alfv\'en loci, around $z/H=1.5$. These are the sites of mode couplings investigated in detail by \citet{SchCal06aa} and by \citet{CalHan11aa} respectively. The combination of all three of these mechanisms, together with the fast wave reflection seen near $z/H=3$, provides a rich array of behaviours for MHD wave propagation upward through magneto\-atmospheres.

Alfv\'en waves \citep{Alf42aa} are archetypally incompressive ($\Div\boldv=0$, or $\bk\,\vdot\,\boldv=0$ in terms of the wave vector $\bk$) and transverse ($\B\,\vdot\,\boldv=0$) in a uniform magneto\-hydro\-dynamic (MHD) plasma, where $\B$ is the magnetic field and $\boldv$ is the plasma velocity. Such `pure' Alfv\'en waves can persist even if the plasma is not uniform, provided there is an available transverse direction that is ignorable. There are two common examples. The first is inclined magnetic field in the $x$-$z$ plane, independent of $y$, with density stratification $\rho(z)$ in the $z$ direction, where pure shear Alfv\'en waves polarized in the $y$-direction are decoupled from any acoustic or gravitational influence. A second example is torsional Alfv\'en waves in vertical axisymmetric magnetic flux tubes, where the polarization is in the azimuthal direction $\theta$.

We now generalize the first case by orienting a uniform magnetic field out of the $x$-$z$ plane in which the wave vector lies. In this case, $\B\vcross\bk$ has a $z$-component, and so an oscillating plasma element moving perpendicular to both $\B$ and $\bk$ must access regions of varying density, and hence of varying Alfv\'en speed $a=B/\sqrt{\mu\rho}$. Therefore it would no longer be decoupled from magneto-acoustic effects.

However, in a very high-$\beta$ (weak field) plasma, the Alfv\'enic wavelength $2\pi/\kpar$ is very short compared to the gravitational scale height $H$, in which case ($\kpar H\gg1$) the atmosphere may be effectively assumed uniform, in the eikonal sense, and an effectively pure Alfv\'en wave may exist. But as this propagates upwards along the field lines, $\kpar H\gg1$ eventually fails and coupling must occur.

Alfv\'en waves generated near the solar photosphere are commonly invoked to explain heating of the solar atmosphere and acceleration of the solar wind \citep{Cravan05aa,TomMcIKei07aa,De-McICar07aa,JesMatErd09aa,McIDe-12aa,MatJesErd13aa,SriBalCal21vr}. Typically, side-to-side shaking of field lines at the photosphere is thought to generate Alfv\'en waves. However, it can also generate slow waves, depending on polarization. Furthermore, these Alfv\'en and slow waves can interconvert between each other in the low chromosphere, so whether any particular oscillation ends up being an Alfv\'en wave or not high in the atmosphere is a complex issue involving the multifarious mechanisms alluded to in Figure \ref{fig:disp}. This is not an academic point. Reference to the dispersion figure shows that only the slow and Alfv\'en waves may reach the upper levels. The slow wave is essentially acoustic at these heights $a\gg c$, and so is prone to shocks and dumping its energy in the chromosphere \citep{NarUlm96aa}. The fast wave, which is predominantly magnetic in $a\gg c$, reflects and typically does not reach the transition region. It is therefore important to understand the routes by which photospheric oscillations may reach the upper atmosphere as Alfv\'en waves, as these are the only viable carriers of oscillation energy into the corona in a plane-stratified atmosphere.

Our discussion does not address coronal loops though. These may act as waveguides, and allow other wave types, kink and sausage for example, to propagate to coronal heights and beyond.

\section{Matrix Equation} \label{sec:mateqn}

Consider a gravitationally stratified plane-parallel isothermal ideal MHD atmosphere permeated by a uniform magnetic field
\begin{equation}\label{B0}
\B_0 = B_0 \left(\sin\theta\cos\phi,\,\sin\theta\sin\phi,\,\cos\theta\right),\end{equation}
directed at angle $\theta$ from the vertical and angle $\phi$ out of the $x$-$z$ plane. It may be assumed that $0\leq \theta < \pi/2$. and $-\pi<\phi\leq\pi$. The atmosphere is characterized by uniform sound speed $c$, density scale height $H$, and exponentially increasing Alfv\'en speed $a=a_0\exp(z/2H)$. 

The linearized perturbed wave equations are expressed in terms of the plasma displacement vector $\bxi=(\xi,\eta,\zeta)$, and a single Fourier mode with horizontal ($x$) and time ($t$) dependence $\exp[i(k x-\omega t)]$ is examined. Without loss of generality, there is no assumed $y$ dependence. The vertical ($z$) behaviour is more complex, and to be found.

Following \cite{CalGoo08aa}, we adopt the dimensionless frequency $\nu=\omega H/c$ and wavenumber $\kappa=k\,H$. The independent variable  $s=\omega H/a=\nu\exp(-z/2H)$ is used instead of $z$, and ranges from $s=0$ at $z=+\infty$ to $s=\infty$ at $z=-\infty$, i.e., it increases downward. We have arbitrarily placed $z=0$ at the equipartition level where $a=c$. In these units, the acoustic cutoff frequency $\omega_c=c/(2H)$ corresponds to $\nu=\half$, and the Alfv\'en-acoustic equipartition level is at $s=\nu$.

The MHD wave equations can be represented as a set of three coupled second-order ODEs in the displacements $\xi(s)$, $\eta(s)$, and $\zeta(s)$. These are conveniently written as a sixth-order matrix equation:
\begin{equation}
s\, \U' = A \,\U,   \label{mateqn}
\end{equation}
where $\U(s)=(\xi,\eta,\zeta,s\xi',s\eta',s\zeta')^T=(\bxi,s\,\bxi')^T$ and
the coefficient matrix $A(s)=A_0+A_2 s^2$ is quadratic in $s$, with both $A_0$ and $A_2$ constant. $A_2$ is of rank 2 and is nilpotent of order 2, i.e., $A_2^2=0$. The matrix $A$ is set out in detail in Appendix \ref{sec:mat}.

The point $s=0$ (i.e., $z=\infty$) is a regular singular point of the equation, and hence is amenable to Frobenius solution \citep{CalGoo08aa}. Details are set out in Appendix \ref{sec:frob}. On the other hand, $s=\infty$ is an irregular singular point, and requires asymptotic solution (Appendix \ref{sec:botAs}). Complete asymptotic expansions for each of the slow, Alfv\'en and fast waves as $s\to\infty$ (i.e., $z\to-\infty$) are presented for the first time, and are easily implemented to deliver excellent accuracy.

\section{Phase Space Insights}  \label{sec:phase}

In terms of our dimensionless variables, the dispersion relation of \citet{NewCal10aa} takes the form
\begin{equation} \label{dispreln}
\left(s^2-\kappa_\parallel^2\right)
\left[s^4-\frac{s^2}{\nu^2}(s^2+\nu^2)(\kappa^2 + K^2)+\frac{s^2}{\nu^2}(\kappa^2+K^2)\kappa_\parallel^2 +\frac{s^4}{\nu^4} n_{\rm BV}^2 \kappa^2 
-\quart\left( s^2-(\kappa^2+K^2)\cos^2\!\theta\right)\frac{s^2}{\nu^2} \right]
 +\frac{s^4\kappa^2}{4\nu^2}\sin^2\!\theta\sin^2\!\phi = 0,
\end{equation}
where $K=k_zH$ is the dimensionless vertical wavenumber, $\kappa_\parallel=\kappa\sin\theta\cos\phi+K\cos\theta$ and $n_{\rm BV}=\sqrt{\gamma-1}/\gamma$ is the dimensionless {\bv} (buoyancy) frequency. Recalling that $s=\omega H/a=\nu \exp(-z/2H)$ is a scaled frequency at fixed height, but also a monotonic function of height at fixed frequency, this dispersion relation defines loci in $z$--$k_z$ space at fixed $\nu$ and $\kappa$. In the eikonal sense \citep{Wei62aa}, the vertical wavenumber $k_z$ may therefore be regarded as a function of height. We shall not use equation (\ref{dispreln}) in our calculations, but it is invaluable in interpreting results. 

Figure \ref{fig:disp} displays two $z$--$k_z$ dispersion diagrams exhibiting the loci of slow, Alfv\'en and fast waves corresponding to mildly inclined magnetic field $\theta=15^\circ$ and two orientations $\phi=15^\circ$ and $\phi=165^\circ$. In the former, the wave with $k_z>0$ is propagating `with the grain' of field inclination, and in the latter it is going `against the grain'. 

Mode coupling is an essentially resonant process \citep{CalAnd10aa}. It \emph{may} occur only where two erstwhile independent waves are immediately adjacent in phase space. Several such locations are evident in figure \ref{fig:disp}, notably
\begin{itemize}
\item fast-slow coupling near $a=c$ ($z=0$) and $k_z>0$ in the left frame, where upgoing fast and slow waves can exchange energy as they propagate `with the grain' \citep{SchCal06aa}; weaker for downgoing $k_z<0$ which are `against the grain';
\item fast-slow coupling near $a=c$ ($z=0$) and $k_z<0$ in the right frame, where downgoing fast and slow waves can exchange energy as they propagate `with the grain', which is now directed downward; weaker for upgoing $k_z>0$;
\item fast-Alfv\'en coupling near $z/H=1.5$, $k_z>0$, again `with the grain' for upgoing waves in the left frame, but weaker for downgoing \citep{CalHan11aa};
\item fast-Alfv\'en coupling near $z/H=1.5$, $k_z<0$, now `with the grain' for downgoing waves in the right frame, but weaker for upgoing;
\item slow-Alfv\'en over extended regions in $z<0$ for both upgoing and downgoing waves.
\end{itemize}
It is this last concurrence that provides the opportunity for mode coupling between slow and Alfv\'en waves that is the primary issue at hand here, though the other processes are also found to play important roles. In particular, note that any \emph{reflection}, which will be seen regularly in our results, can only occur via the intermediary of the reflecting fast wave, seen in the diagrams at $z/H\approx3$.

However, an opportunity for mode coupling is not sufficient, a mechanism is also required. Neither fast-Alfv\'en nor slow-Alfv\'en couplings actually occur in 2D ($\sin\phi=0$) in ideal MHD, though fast-slow remains. Three-dimensionality is required to engage the stratification, which removes the ignorable direction of Alfv\'en polarization.

\section{Numerical Survey}
Consider a pure Alfv\'en or slow wave of dimensionless frequency $\nu$ and dimensionless $x$-wavenumber $\kappa$ injected at $s\gg \max(\nu,\ \kappa)$ in an isothermal plane stratified atmosphere with magnetic field $\B_0$ specified by equation (\ref{B0}). Fast-slow mode conversion happens at $s=\nu$ and fast wave reflection\footnote{Neglecting the acoustic cutoff and {\bv} frequencies, reflection occurs at $s^2=\kappa^2(\nu^2-\kappa^2\sin^2\theta\cos^2\phi)/(\nu^2-\kappa^2)$, or roughly $s\approx\kappa$ if $\nu\gg\kappa$.} near $s=\kappa$ (assuming $\kappa\ll\nu$). Fast Alfv\'en coupling occurs in 3D ($\sin\phi\ne0$) in the vicinity of the fast reflection height \citep{CalHan11aa}. Hence $s\gg \max(\nu,\ \kappa)$ puts us well below all interaction regions. The ratio of specific heats $\gamma$ is set to $\fract{5}{3}$ throughout. 

The six imposed boundary conditions are that there be no incoming (downgoing) waves at the top $s\to0$, and only a single injected Alfv\'en or slow wave at the bottom $s\to\infty$ which carries unit wave energy flux. Outgoing waves of all three types are allowed at each boundary. We then calculate the total transmitted slow (acoustic-gravity) flux $F_s^{_\uparrow}$ and Alfv\'en $F_A^{_\uparrow}$ fluxes at the top, and the downward slow (magnetic) flux $F_s^{_\downarrow}$, Alfv\'en flux $F_A^{_\downarrow}$ and fast (acoustic-gravity) flux $F_f^{_\downarrow}$ at the bottom. By energy conservation, $F_s^{_\uparrow}+F_A^{_\uparrow}-F_s^{_\downarrow}-F_A^{_\downarrow}-F_f^{_\downarrow}=1$. There is no top fast wave flux, as the fast wave is necessarily evanescent (for $\kappa\ne0$).

Slow and Alfv\'en injected waves at the base are distinguished by their polarization. In the $s\to\infty$ limit, both waves are identically vertical since the vertical wavenumber $k_z\to\infty$. As discussed in Appendix \ref{sec: mag}, this orients the slow wave displacement vector in the direction $(\cos\phi,\,\sin\phi,\,0)$ and the Alfv\'en wave has perpendicular polarization $(-\sin\phi,\,\cos\phi,\,0)$. 

Complete asymptotic expansions for each of these two cases are developed in Appendix \ref{sec: mag}, and if employed at sufficiently large matching point $s_m$ ($s_m=10$ is typically sufficient at small to moderate $\theta$ with the parameters we consider), yield very high accuracy approximations using the optimal truncation rule. A similar complete asymptotic solution is presented in Appendix \ref{sec: ag} for the fast wave. Convergent Frobenius series solutions about $s=0$ (Appendix \ref{sec:frob}) are easily applicable at these values of $s$, so no numerical integration is required.

\subsection{Variation with Magnetic Field Orientation $\phi$}

\begin{figure*}
\begin{center}
\includegraphics[width=\textwidth]{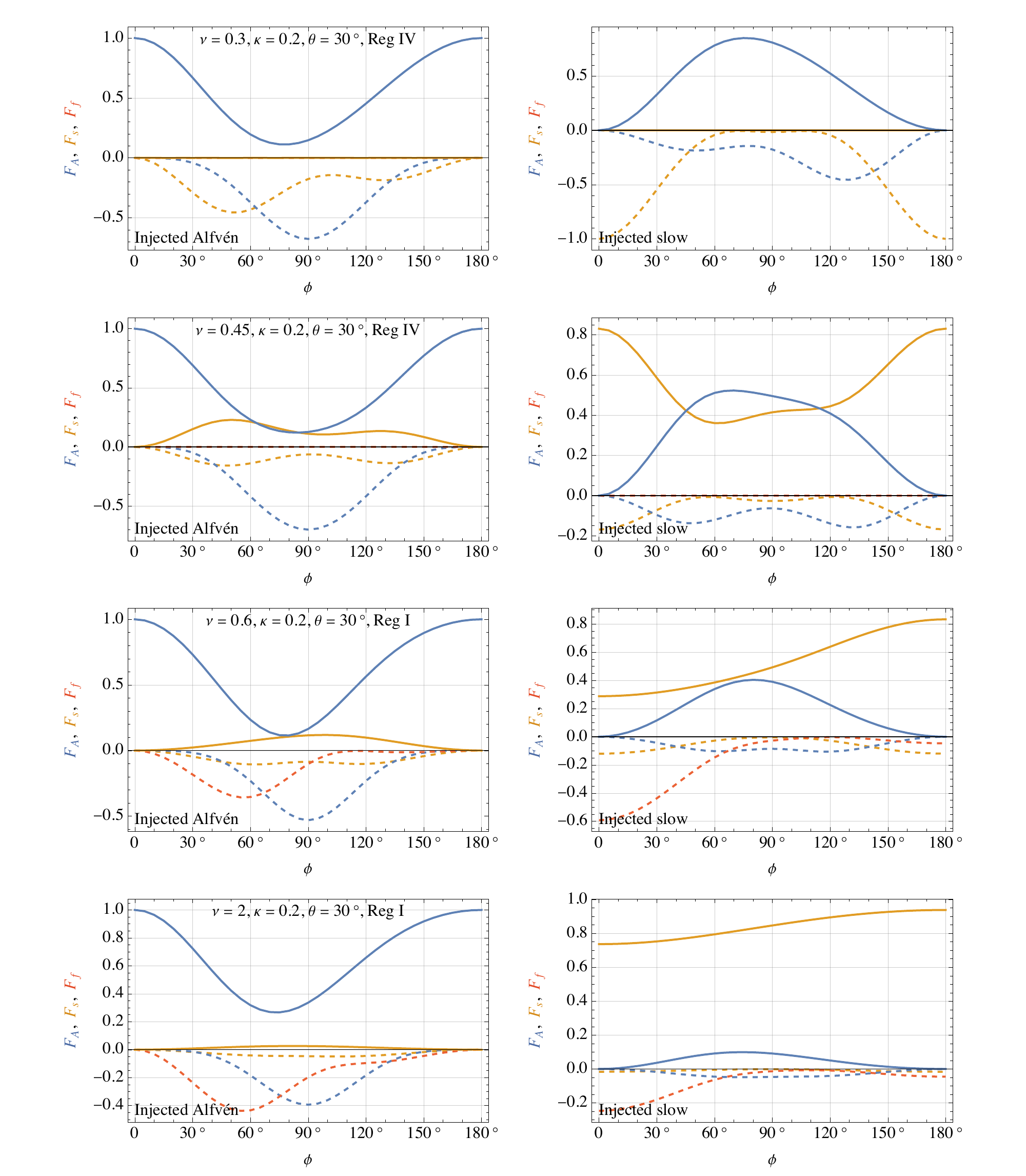}
\caption{Escaping vertical wave energy fluxes as functions of angle $\phi$ for inclined magnetic field $\theta=30^\circ$ and dimensionless $x$-wavenumber $\kappa=0.2$ at various dimensionless frequencies $\nu=0.3$ (top row), 0.45 (second row), 0.6 (third row) and 2 (bottom), in the cases of injected Alfv\'en wave at the bottom (left column) and injected slow wave (right column). The full blue curve represents the Alfv\'en flux $F_A^{_\uparrow}$ escaping upward at the top and the full orange curve is the slow (acoustic) wave flux $F_s^{_\uparrow}$ at the top. The dashed curves represent escaping (downward) fluxes at the bottom: blue -- Alfv\'en $F_A^{_\downarrow}$; orange -- slow (magnetic) $F_s^{_\downarrow}$; and red -- fast (acoustic) $F_f^{_\downarrow}$.}
\label{fig:nuGrid30}
\end{center}
\end{figure*}

Figure \ref{fig:nuGrid30} plots $F_s^{_\uparrow}$, $F_A^{_\uparrow}$, $F_s^{_\downarrow}$, $F_A^{_\downarrow}$ and $F_f^{_\downarrow}$ against $\phi$ in the case of inclined magnetic field, $\theta=30^\circ$ for injected Alfv\'en and slow waves of various frequencies. Recalling that $\nu=0.5$ corresponds to the acoustic cutoff frequency, which is around 5 mHz in the solar chromosphere, our selected frequencies $\nu=0.3$, 0.45, 0.6 and 2 are roughly 3 mHz, 4.5 mHz, 6 mHz and 20 mHz respectively.  We fix the dimensionless horizontal wavenumber $\kappa=0.2$ throughout, corresponding to $k\approx2$ $\rm rad\ Mm^{-1}$ if $H\approx100$ km. Reducing wavenumber to $\kappa=0.05$ produces similar results (not shown). Increasing it to $\kappa=0.6$ strengthens slow-to-Alfv\'en production.

As expected, $F_A^{_\uparrow}=1$ for the injected Alfv\'en wave at $\phi=0^\circ$ and $180^\circ$, since in that case the Alfv\'en wave is decoupled from the magneto\-acoustic waves. In all cases though, the Alfv\'en flux reaching the top is severely diminished as $\phi$ increases toward $90^\circ$. The bulk of the missing upward Alfv\'en flux near $\phi=90^\circ$ reappears as downgoing Alfv\'en flux at the bottom, representing an effective reflection process. This can only have occurred via a multi-stage process, Alfv\'en-to-slow-to-fast-reflected-to-slow-to-Alfv\'en.

Conversely, injected slow waves near $\phi=90^\circ$ convert significantly but not totally to Alfv\'en waves at the top, though this reduces with increasing frequency. 

At $\nu=0.3$ we are in the evanescent Region IV of the acoustic-gravity propagation diagram Figure \ref{fig:PropDiag}, so $F_f^{_\downarrow}=0$ as the fast wave does not propagate vertically in the high-$\beta$ region. Similarly, since $\nu<\half\cos\theta$, the ramp frequency, $F_s^{_\uparrow}=0$ also. The ramp effect operates in low-$\beta$ plasma with inclined magnetic field to reduce the acoustic cutoff frequency by the factor $\cos\theta$ \citep{BelLer77aa,JefMcIArm06aa}. The ramp effect is very evident in the fast-wave Frobenius eigenvalues, equation (\ref{mu}), through the terms $\sqrt{4\nu^2-\cos^2\theta}$, which correspond to travelling waves only if $\nu>\half\cos\theta$.

At $\nu=0.45$, though we are still in Region IV with $F_f^{_\downarrow}=0$, the frequency now exceeds the ramp value $0.433$, and considerable slow (acoustic) flux $F_s^{_\uparrow}$ is generated at the top, especially from the injected slow wave. This is caused by slow-to-slow mode transmission at the Alfv\'en-acoustic equipartition height $a=c$, i.e., $s=\nu$, \citep{SchCal06aa}. The weaker generation of $F_s^{_\uparrow}$ from the injected Alfv\'en wave is due to a combination of Alfv\'en-to-slow conversion in the high-$\beta$ region followed by slow-to-slow transmission through $a=c$, and Alfv\'en-to-slow conversion followed by slow-to-fast conversion at $a=c$ followed by fast-to-Alfv\'en conversion near $\omega=a k$, i.e., $s=\kappa$ \citep{SriBalCal21vr}.

At $\nu=0.6$ we are above the acoustic cutoff frequency, so both $F_f^{_\downarrow}$ and $F_s^{_\uparrow}$ can be non-zero. The latter is particularly prominent for the injected slow wave case, especially at $\phi\approx180^\circ$. This is to be expected, as then the wave is propagating `against the grain' (the magnetic field is inclined contrary to the positive-$x$ propagation direction), which is the most favoured scenario for fast-to-fast (i.e., magnetic-to-acoustic) mode conversion \citep{SchCal06aa}. This case continues to produce substantial conversion of an injected Alfv\'en wave, and some conversion to an Alfv\'en wave from an injected slow wave.

Finally, at much higher frequency $\nu=2$, Alfv\'en-to-slow and vice versa conversion are both weakened. This is to be expected as higher-frequency shorter-wavelength waves should not couple as readily. Again, the direct slow-to-slow conversion at $a=c$ is prominent, and favoured at $\phi\approx180^\circ$.

\begin{figure*}
\begin{center}
\includegraphics[width=\textwidth]{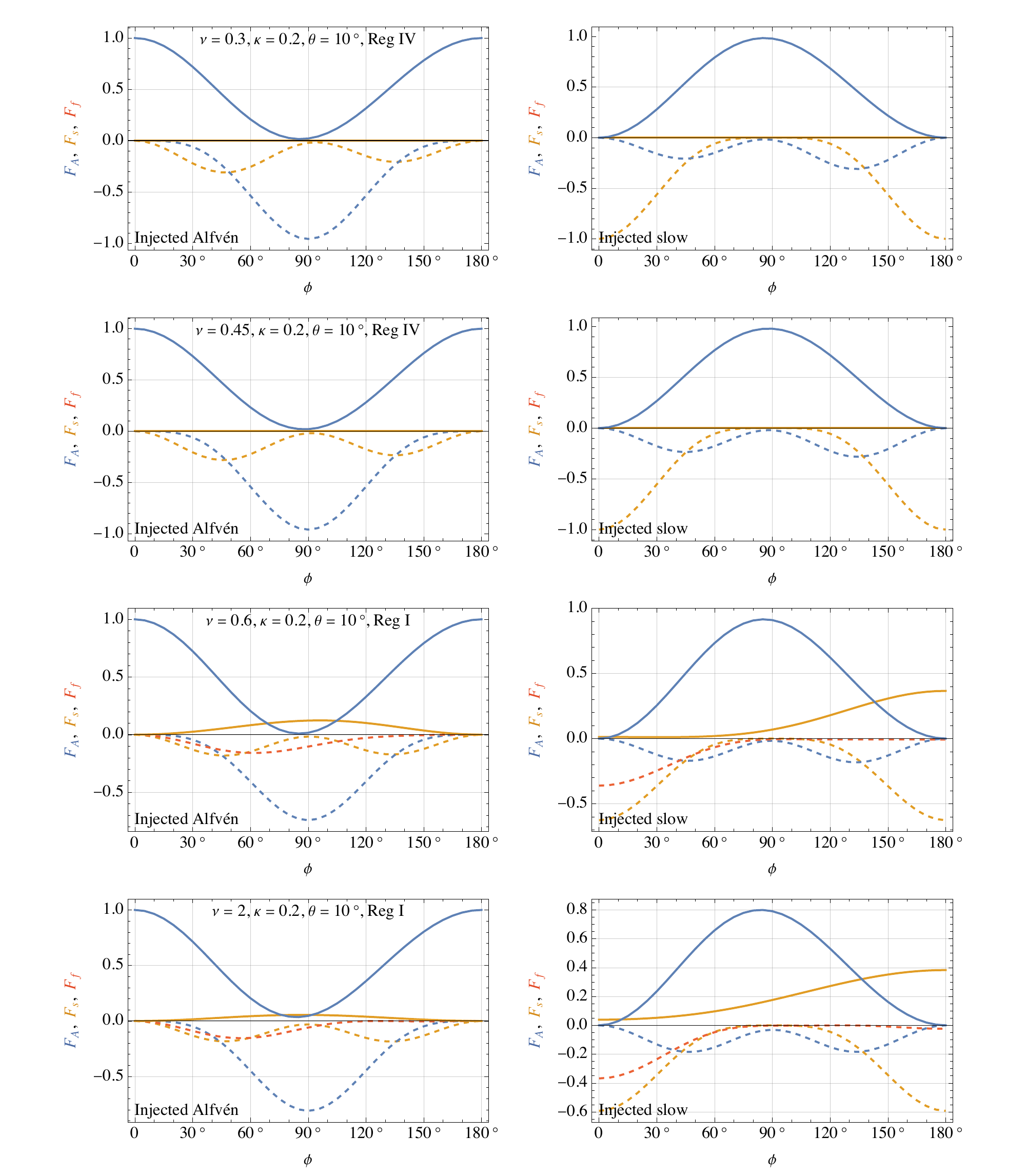}
\caption{Same as Figure \ref{fig:nuGrid30}, but for $\theta=10^\circ$.}
\label{fig:nuGrid10}
\end{center}
\end{figure*}

With lesser field inclination, $\theta=10^\circ$, Figure \ref{fig:nuGrid10} shows similar behaviour, except that
\begin{enumerate}
\item the ramp effect barely operates ($\half\cos10^\circ=0.492$), so there is no acoustic flux at either top or bottom; and
\item the conversion of the injected Alfv\'en wave and the injected slow wave are now near-total at $\phi\approx90^\circ$, especially at lower frequencies. 
\end{enumerate}

\begin{figure*}
\begin{center}
\includegraphics[width=\textwidth]{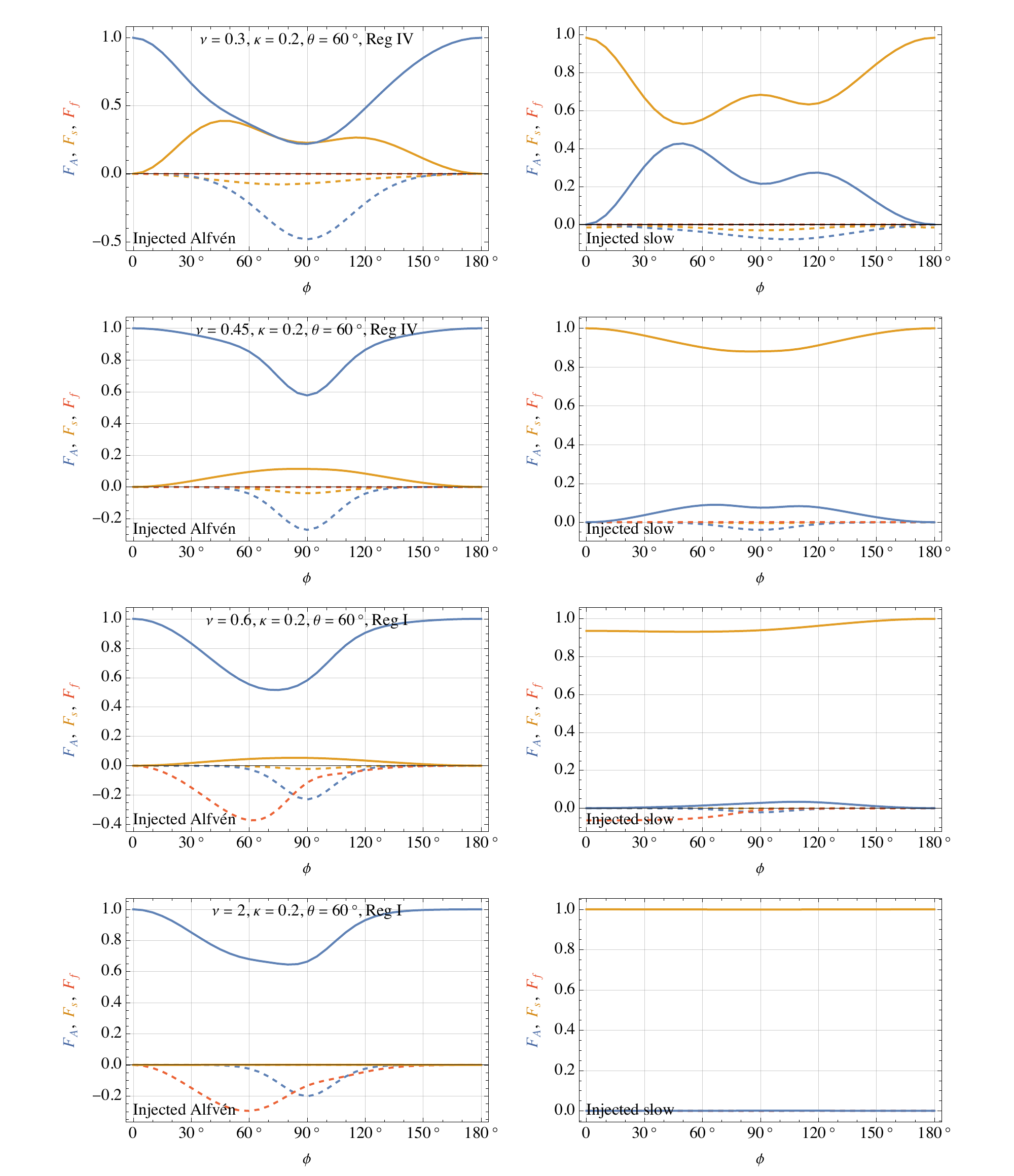}
\caption{Same as Figure \ref{fig:nuGrid30}, but for $\theta=60^\circ$.}
\label{fig:nuGrid60}
\end{center}
\end{figure*}

At $\theta=60^\circ$ (Figure \ref{fig:nuGrid60}), direct Alfv\'en wave penetration is significantly enhanced, and slow-to-Alfv\'en conversion reduced. Indeed, the slow wave is largely just converted to the slow (acoustic) wave at the top, via the mechanism of \citet{SchCal06aa}, since the attack angle of the wavevector to the magnetic field at $a=c$ is large.

\subsection{Fluxes averaged over $\phi$}  \label{sec:Av}
\begin{figure*}
\begin{center}
\includegraphics[width=\textwidth]{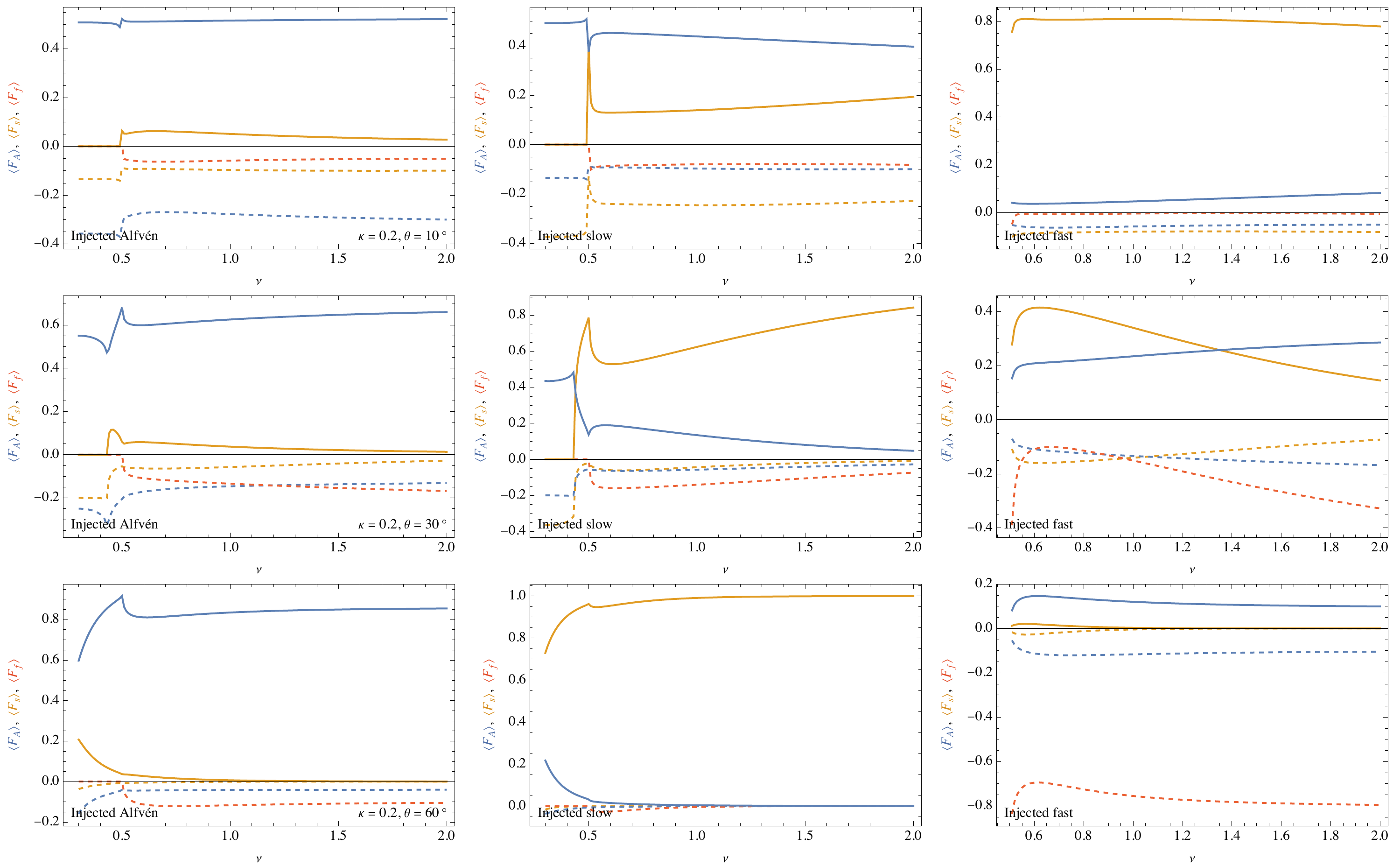}
\caption{Top and bottom fluxes $\langle F_A^{_\uparrow}\rangle$ etc., averaged over all $\phi$, for the cases of Figures \ref{fig:nuGrid10}, \ref{fig:nuGrid30} and \ref{fig:nuGrid60}, i.e., $\kappa=0.2$ with $\theta=10^\circ$ (top row), $30^\circ$ (middle row) and $60^\circ$ (bottom row), plotted as functions of frequency $\nu$. Left column: injected Alfv\'en wave; centre column: injected slow wave; right column: injected fast wave.}
\label{fig:Fav}
\end{center}
\end{figure*}

As the directions of wave excitation at the bottom might be expected to be random, we average the various fluxes over all $\phi$, equivalent to fixing the magnetic field orientation and averaging over the direction of the horizontal wavevector $\bk$, for $\kappa=0.2$ and $\theta=10^\circ$, $30^\circ$ and $60^\circ$, at a range of frequencies from 0.3 to 2 (Figure \ref{fig:Fav}).

It is seen that $\langle F_A^{_\uparrow}\rangle$ ranges from about 0.5 at all frequencies for $\theta=10^\circ$ up to about 0.8 at $60^\circ$ for the injected Alfv\'en wave, where $\langle\ldots\rangle=\pi^{-1}\int_0^\pi \ldots\,d\phi$. Simultaneously, the reflected Alfv\'en wave $\langle F_A^{_\downarrow}\rangle$ wains with increasing field inclination. 

However, for the injected slow wave, $\langle F_A^{_\uparrow}\rangle$ falls from about 0.4 at $\theta=10^\circ$ to become negligible at larger $\theta$ and especially at larger $\nu$. $\langle F_A^{_\downarrow}\rangle$ also diminishes with increasing $\theta$. This appears to be because the bulk of the flux goes into $\langle F_s^{_\uparrow}\rangle$ in those cases.

The sharp features near $\nu=0.5$ are due to the acoustic cutoff effect, combined with the ramp effect at the top.

For completeness and comparison, the case of fast injection has been included (above $\nu=0.51$ so that it propagates), as the third column, despite it not being the topic of this article. The small-to-moderate amounts of top Alfv\'en flux generated are due to fast-to-fast (i.e., acoustic-to-magnetic) conversion at $a=c$ followed by fast-to-Alfv\'en 3D conversion near the fast reflection height, so no Alfv\'en flux is produced at $\phi=0^\circ$ or $180^\circ$. This process was addressed in \citet{CalGoo08aa}. Fast wave injection will not be discussed further.

\section{Conclusions and Discussion}

We have presented two major results. First, the complete asymptotic solutions as $s\to\infty$ of the $6^{\rm th}$ order wave equations for an isothermal plane stratified magneto\-atmosphere with uniform arbitrarily directed magnetic field are new, and complete the leading order results given by \citet{ZhuDzh84aa}, equation (11). They are easily coded and computationally practical. In most circumstances, a combination of the asymptotic and Frobenius solutions is sufficient to cover all $s$ to good accuracy without need of numerical integration. 

Second, we have demonstrated that generation of 3D Alfv\'en waves of arbitrary orientation with respect to the inclined magnetic field direction only couples strongly to high-atmosphere upward-travelling Alfv\'en waves for small $\sin\phi$, and indeed is weakly coupled at $\phi\approx90^\circ$, especially at small inclination angles $\theta$. This phenomenon is associated with a peak in reflected Alfv\'en waves, which judging by Figure \ref{fig:disp} can only happen via a multi-stage process including fast wave reflection.

Conversely, injected slow waves at $\phi\approx90^\circ$ are strongly converted to upgoing Alfv\'en waves at the top, and also to slow (acoustic) waves at significant $\theta$.

In reality, transverse wave driving by photospheric granulation will be random in direction, and similarly for its Fourier decomposition into horizontal wavevector $\bk$. Correlation between the direction of polarization and that of $\bk$ depends on characteristics of the granulation, and is beyond our scope here, so the fractions of energy in the injected slow and Alfv\'en waves is uncertain.

Nevertheless, with $\bk$ arbitrarily fixed in the $x$-direction we have calculated the various resulting fluxes independently averaged over angle $\phi$. It is found that the fraction of injected Alfv\'en flux reaching the top as an Alfv\'en wave increases from around 50\% at small field inclination $\theta$ to 80\% at $\theta=60^\circ$. On the other hand, the averaged Alfv\'en flux resulting from injected slow waves diminishes rapidly with increasing $\theta$ and $\nu$, largely due to the slow flux $\langle F_s^{_\uparrow}\rangle$ taking most of the power. The two results are complimentary: as field inclination and frequency increase, slow-Alfv\'en coupling along their near-coincident dispersion curve loci weakens, leaving traditional fast-slow mode conversion to dominate the injected slow wave.

A take-home message from this study is that Alfv\'en flux generated at the photosphere in an ideal MHD stratified medium is only about 50\% effective in reaching the upper atmosphere as Alfv\'en flux in near-vertical magnetic field, but becomes more effective as field inclination increases. This ignores the Hall effect and the transition region, but provides a baseline against which models with those features can be measured. Conversely, generated slow waves can significantly pass through as Alfv\'en waves at small $\theta$, but become ineffective at moderate to large inclinations.


\section*{Data Availability}
The data underlying this article will be shared on reasonable request to the corresponding author.



\bibliographystyle{mnras}
\bibliography{fred} 



\appendix

\section{Matrix wave equation}  \label{sec:mat}
The component of the matrix $A(s)$ appearing in equation (\ref{mateqn}) are set out in \citet{CalGoo08aa}, but are repeated here for completeness:
\begin{equation}
\A(s)=
\begin{pmatrix}
\mathbfss{0} & \I\\
\P & \Q
\end{pmatrix},
\end{equation}
where each of the four constituent blocks is $3\times3$. Here $I$ is the identity matrix,
\begin{equation}\label{Q}
\Q=
\begin{pmatrix}
 2 \ri \kappa  \cos \phi  \tan \theta  & -2 \ri \kappa  \sin \phi  \tan \theta  & 2 \ri \kappa  \left(\frac{s^2 \sec ^2\theta }{\nu ^2}+\tan ^2\theta \right)-2
   \cos \phi  \tan \theta  \\
 0 & 4 \ri \kappa  \cos \phi  \tan \theta  & -2 \sin \phi \tan \theta  \\
 2 \ri \kappa  & 0 & 2 \ri \kappa  \cos \phi  \tan \theta -2
\end{pmatrix},
\end{equation}
and the components $P_{ij}$ of $\P$ are
\begin{subequations}
\begin{align} \label{P}
P_{11} &=-2 \left(\cos 2 \phi  \tan ^2\theta -1\right) \nu ^2+\left(\left(\frac{4 \kappa ^2}{\nu ^2}-4\right) s^2+4 \kappa ^2-2 \nu ^2\right) \sec ^2\theta 
+4 \ri(1-\gamma^{-1}) \kappa  \cos \phi  \tan \theta \\
P_{12}&=-2 (\kappa ^2+\nu ^2) \sin 2 \phi  \tan ^2\theta\\
P_{13}&=\frac{4 \ri \left(\cos \phi  \tan \theta  \left(i \gamma  \left(\kappa ^2+\nu ^2\right)+\kappa  \cos \phi  \tan \theta \right) \nu ^2+s^2 \kappa  \sec ^2\theta\right)}{\gamma  \nu ^2}\\
P_{21}&=\frac{4 \ri \sin \phi  \tan \theta  \left(i \gamma  \cos \phi  \tan \theta  \nu ^2+(\gamma -1) \kappa \right)}{\gamma }\\
P_{22}&=-4 \left(s^2 \sec ^2\theta +\left(\nu ^2 \sin ^2\phi -\kappa ^2 \cos ^2\phi \right) \tan ^2\theta \right)\\
P_{23}&=-\frac{4 \sin \phi  \tan \theta  \left(\gamma  \nu ^2-\ri \kappa  \cos \phi  \tan \theta \right)}{\gamma }\\
P_{31}&=\frac{4 \ri (\gamma -1) \kappa }{\gamma }+4 (\kappa^2 -\nu^2 ) \cos \phi  \tan \theta \\
P_{32}&=-4 \nu ^2 \sin \phi  \tan \theta \\
P_{33}&=\frac{4 \ri \kappa  \cos \phi  \tan \theta }{\gamma }-4 \nu ^2,
\end{align}
\end{subequations}
where $\gamma$ is the ratio of specific heats. 


\section{Frobenius series solution}  \label{sec:frob}
The point $s\to0$ ($z=\infty$) is a regular singular point of equation~(\ref{mateqn}), and hence supports Frobenius series solution
\begin{equation}
\U(s)=\sum_{n=0}^\infty \boldu_n s^{n+\mu}.  \label {frob}
\end{equation}
The indices $\mu$ are the eigenvalues of $\A_0$,
\begin{equation} \label{mu}
\mu\in
\Bigl\{-2 \kappa ,\,2 \kappa ,\,-\ri \sqrt{4 \nu ^2-\cos ^2\theta }\, \sec \theta +2 \ri \kappa  \cos \phi  \tan \theta -1,
\ri \sqrt{4 \nu ^2-\cos ^2\theta } \,\sec
   \theta +2 
   \ri \kappa  \cos \phi  \tan \theta -1,\,2 \ri \kappa  \cos \phi  \tan \theta ,\,2 \ri \kappa  \cos \phi  \tan \theta \Bigr\},
\end{equation}
corresponding respectively to the growing and evanescent fast modes, the outgoing and incoming slow modes (if $\nu>\half\cos\theta$; otherwise growing and evanescent), and the Alfv\'en mode (double root). The Alfv\'en eigenvalue only has geometric multiplicity 1, and so $A_0$ is not diagonalizable. 

The even $\boldu_n$ coefficients for each of the first five eigen-solutions are found via the recurrence relation
\begin{equation}
\boldu_n=\left((n+\mu)\I-\A_0\right)^{-1}\A_2 \boldu_{n-2} , \label{frobrec}
\end{equation}
where $\boldu_0$ is the eigenvector belonging to $\mu$. All the odd $\boldu_n$ are zero.

This recovers the leading terms of the Frobenius solution of \cite{CalGoo08aa}. As there are no other finite singular points of the differential equation, the series has infinite radius of convergence. Indeed, the full Frobenius series converge rapidly for even quite large $s$, after a strong peak in the coefficients near some $n=n_{\rm m}$ that increases with increasing $s$. The series are therefore quite usable in a practical sense.

The sixth solution has no eigenvector, and instead must be constructed using the generalized eigenvector. This results in a logarithmic term
\begin{equation}
\U_6(s)=\frac{2}{\pi}\left( \U_5(s)\ln s+\sum_{n=0}^\infty\boldv_n s^{n+\mu_6} \right)  .  \label{U6}
\end{equation}
The $\boldv_n$ coefficients are found from the recurrence
\begin{subequations}
\begin{align}
&\left(\A_0-\mu_6\I\right)\boldv_0 =\boldu_0 ,\\[4pt]
&\boldv_1 = \bf{0} ,\\[4pt]
&\left(\A_0-(n+\mu_6))\I\right)\boldv_n =\boldu_n-\A_2\boldv_{n-2} ,
\end{align}
\end{subequations}
where the $\boldu_n$ belong to $\U_5$.

Both $\U_5$ and $\U_6$ correspond to standing Alfv\'en waves. Upward $\U_+$ and downward $\U_-$ propagating Alfv\'en waves may be constructed as
$\U_\pm=(1\mp i \alpha)\U_5\mp i\, \U_6$ for arbitrary real $\alpha$. The wave energy flux carried by this wave is independent of $\alpha$.

Typically, in our numerical calculations matching to asymptotic series at $s\approx10$, the terms $\|\boldu_n\| s^n$ increase to a strong maximum at $n$ of a few tens, but then decrease rapidly to deliver very fast exponential convergence thereafter.

\section{Bottom Asymptotics}  \label{sec:botAs}
Unlike $s=0$, the bottom of the domain ($s=\infty$) is an \emph{irregular} singular point of the differential equations. A convergent series solution is therefore not available. Complete asymptotic expansions may be derived though. The fast wave is an acoustic-gravity wave to leading order. The magnetic waves are pure slow and Alfv\'en waves in the $s\to\infty$ limit.

%

\subsection{Acoustic-gravity waves}   \label{sec: ag}
We seek solutions of the form
\begin{equation}
\U(s) \sim \sum_{n=0}^\infty \boldu_n s^{r-n} \mbox{ as $s\to\infty$}.   \label{fastU}
\end{equation}
Substituting this into the matrix differential equation (\ref{mateqn}) and equating coefficients of $s$ we find
\begin{subequations}
\begin{align}
&\A_2 \boldu_0 =0  \label{u0F} \\[4pt]
&\A_2 \boldu_1 =0  \label{u1F}\\[4pt]
&\A_2 \boldu_{n+2} = \left((r-n)\I-\A_0\right)\boldu_n \equiv \matB_n \boldu_n \label{unF}
\end{align}
\end{subequations}
Equation (\ref{u0F}) means that $\boldu_0$ is in the nullspace of $\A_2$. The odd and even coefficients are decoupled, and we may set $\boldu_1=0$ without loss of generality, as it would simply recover the solution based on $\boldu_0$ but with $r$ increased by 1. Hence, all the odd coefficients vanish.

The $6\times6$ matrix $\A_2$ is of rank 2, so the recurrence is singular and awkward to apply: each iteration yields a $\boldu_{n+2}$ that contains four arbitrary constants that can only be determined by requiring the right hand sides of later iterations to be in the column space of $\A_2$. However, the process can be rendered in more convenient form by introducing $\L=\diag[1,1,1,0,0,1]$ and 
\begin{equation}
\G_n = \A_2+\L \matB_{n+2}.   \label{G}
\end{equation}
Noting that $\L\A_2=0$ and so $\L\matB_{n+2}\boldu_{n+2}=0$, the recurrence relation can be rewritten as
\begin{equation}
\G_n \boldu_{n+2}= \matB_n\boldu_n,   \label{Grec}
\end{equation}
where $\G_n$ is non-singular for $n\ge-1$.

However, $\G_{-2}$ must be singular to obtain a nontrivial solution. This provides the indicial equation by which $r$ may be determined by setting $\det \G_{-2}=0$:
\begin{equation}
16\sec^4\theta\left((r+1)^2+4\kappa_z^2\right)=0,  \label{indicial}
\end{equation}
whence
\begin{equation}
r=-1 + 2i  \kappa_z,  \label{rF}
\end{equation}
where 
\begin{equation}
\kappa_z=\pm\left(\nu^2-\kappa^2+\frac{\gamma-1}{\gamma^2}\frac{\kappa^2}{\nu^2}-\quart\right)^{1/2}
\end{equation}
is the acoustic-gravity dimensionless vertical wavenumber. 

The $+$ sign in $\kappa_z$ corresponds to a wave with downgoing phase, whilst the $-$ sign is upgoing phase. However, the vertical components of phase and group velocity in the gravity wave regime are oppositely directed. To select the correct \emph{outward} acoustic gravity branch at the bottom, we should choose the appropriate sign for each of the acoustic-gravity propagation diagram regions:
\begin{description}
  \item[\bf Region I:] $+$ sign in the travelling acoustic wave region $\kappa_z^2>0$, $\nu\ge\half$ (above the acoustic cutoff frequency);
  \item[\bf Region II:] $-$ sign in the travelling gravity wave region $\kappa_z^2>0$, $\nu^2<n_{\rm BV}^2=(\gamma-1)/\gamma^2$ (below the {\bv} frequency);
  \item[\bf Regions III and IV:] $+$ sign in the evanescent regions $\kappa_z^2<0$ to make $\kappa_z$ positive imaginary;
\end{description}
and the reverse for the inward wave.

\begin{figure}
\begin{center}
\includegraphics[width=.35\textwidth]{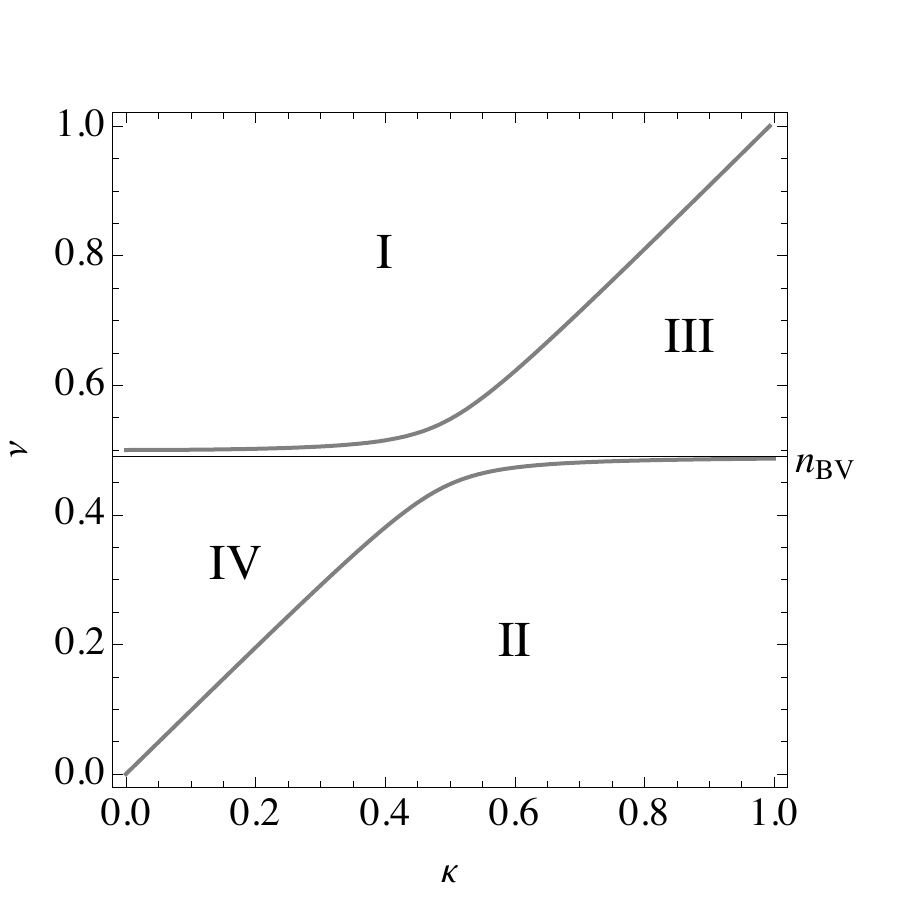}
\caption{Acoustic-gravity propagation diagram. Region I corresponds to vertically propagating acoustic waves, Region II to propagating gravity waves, and Regions III and IV to evanescent waves.}
\label{fig:PropDiag}
\end{center}
\end{figure}

With a convenient normalization,
\begin{equation}  \label{u0}
\boldu_0 =\left(-\frac{i}{2} \kappa  \left(\gamma  \left(2i \kappa _z-1\right)+2 \right),\ 0,\,\gamma  \left(\kappa
   ^2-\nu ^2\right),\,-\frac{i}{2} \kappa  \left(2i \kappa _z-1\right) \left(2+\gamma  \left(-1+2 i \kappa
   _z\right)\right),\,0,\, \gamma  \left(\kappa ^2-\nu ^2\right) \left(2 i\kappa _z-1\right)\right)^T.
\end{equation}
This leading term contains no reference to the magnetic field ($\theta$ and $\phi$ specifically), and is identical to the acoustic-gravity wave in a nonmagnetic atmosphere.

Subsequent coefficients $\boldu_n$, $n=2,\,4,\, \ldots$ may be calculated analytically, but are algebraically very long and complicated. In practice, it is easier to calculate these numerically using equation (\ref{Grec}) with given $\nu$, $\kappa$ etc.

\subsection{Magnetic waves}  \label{sec: mag}
Guided by the known exact solution for the 2D case \citep{Cal09aa}, we seek solutions of the form
\begin{equation}
\U(s) \sim e^{i\,\alpha\,s} \sum_{n=0}^\infty \boldv_n s^{r-n} \mbox{ as $s\to\infty$},   \label{slowU}
\end{equation}
where $\alpha\ne0$ and $r$ are to be determined. This leads to the recurrence
\begin{subequations} \label{vrec}
\begin{align}
&\A_2 \boldv_0 =0  \label{v0S} \\[4pt]
&\A_2 \boldv_1 =i\,\alpha\,\boldv_0  \label{v1S}\\[4pt]
&\A_2 \boldv_{n+2} = \left((r-n)\I-\A_0\right)\boldv_n+i\,\alpha\,\boldv_{n+1}=\matB_n\boldv_n+i\alpha\,\boldv_{n+1}.  \label{vnS}
\end{align}
\end{subequations}
Both odd and even terms are present.

Equations (\ref{v0S}) and (\ref{v1S}) require that $\boldv_0$ be in both the nullspace of $A_2$ and its column space, whence 
\begin{equation}
\boldv_0=(0,0,0,p,q,0)^T,  \label{v0}
\end{equation} 
with $p$ and $q$ arbitrary. The physical interpretation of this result is that both Alfv\'en and slow waves are asymptotically vertical as $s\to\infty$ (i.e., $k_z\gg k$) and so the polarization of both is horizontal. The $\xi$ and $\eta$ displacements do not enter at this stage since they are one lower order in $s$ than $s \xi'\sim i\alpha s\xi$ and $s \eta'\sim i\alpha s\eta$.

This allows us to distinguish the Alfv\'en and slow waves. Since the wavevector is vertical to leading order and the magnetic field is oriented in the $ \left(\sin\theta\cos\phi,\,\sin\theta\sin\phi,\,\cos\theta\right)$ direction, the slow and Alfv\'en wave polarization directions must be respectively in the vertical plane of $\B_0$ and normal to it, i.e., 
\begin{equation}\label{pq}
(p,q) =
  \begin{cases}
    (\cos\phi,\,\sin\phi) & \text{for the slow wave},\\
    (-\sin\phi,\,\cos\phi) & \text{for the Alfv\'en wave}.
  \end{cases}
\end{equation}
For convenience, general $p$ and $q$ are retained and may be substituted as required.

In the vertical field case $\theta=0$ where $\phi$ loses its meaning, it should be assumed that $\phi=0$ for this purpose so that the slow wave is polarized in the $x$-direction and the Alfv\'en wave in the $y$-direction. 

Recall that $\A_2$ only has rank 2, so equations (\ref{vrec}) cannot be used directly to calculate the successive $\boldv_n$ without adding an arbitrary vector from the nullspace. Pre-multiplying equation (\ref{vnS}) by $\A_2$, and recalling that $\A_2$ is nilpotent of degree 2, it follows that
\begin{equation}
\C_n\boldv_n= -i\alpha \matB_{n-1}\boldv_{n-1},  \label{vrecAlt}
\end{equation}
where
\begin{equation} \label{Cn}
\C_n=\A_2\matB_n-\alpha^2 \I.
\end{equation}
Applying this for $n=0$, setting $\boldv_{-1}=0$, shows that $\C_0$ must be singular. Therefore, from the determinant $\alpha^8(\alpha^2-4\sec^2\!\theta)^2$,
\begin{equation}\label{alpha}
\alpha=\pm2\sec\theta, 
\end{equation}
with the $+$ sign corresponding to a downgoing wave and the $-$ sign to upgoing.

To find $r$, we reduce the augmented matrix $(\C_1 | \matB_0 \boldv_0)$ by elementary row operations to
\begin{equation}  \label{Caug}
 \begin{pmatrix}
   1 &0 & 0& 0& 0& 0& -\half i p \cos\theta \\[2pt]
  0  & 1 & 0& 0& 0& 0& -\half i q \cos\theta \\[2pt]
    0 & 0& 1&0 & 0& 0& 0\\[2pt]
    0 & 0&0&0&0&1 & p\kappa\cos\theta\\[2pt]
     0& 0& 0& 0&0 & 0& \half i p\cos\theta \left(r-\half-2i\kappa\cos\phi \tan\theta \right) \\[4pt]
   0  & 0& 0& 0& 0&0 &  \half i q\cos\theta \left(r-\half-2i\kappa\cos\phi \tan\theta \right) \\
 \end{pmatrix},
\end{equation}
which admits solutions only if 
\begin{equation}
r=\half+2i \kappa \cos\phi \tan\theta.  \label{rMag}
\end{equation}

With $\alpha$ and $r$ now determined, all $\C_n$ have rank 4, which is an improvement over the rank 2 matrix $\A_2$ on the left hand side of equations (\ref{vrec}). However, equation (\ref{Cn}) still does not yield a unique solution. It must be supplemented by the requirement that $\matB_{n-1}\boldv_{n-1}\in\colsp(\C_n)$, $n\ge1$, for subsequent coefficients to exist. Letting $\M_n$ be a $6\times6$ matrix whose first four rows are filled with zeros and last two rows form a basis of the nullspace of $\C_n^T$, this may be expressed as $\M_{n+1}\matB_n\boldv_n=0$. Adding this to equation (\ref{Cn}) yields a nonsingular recurrence
\begin{equation}
  \matD_n\boldv_n= -i\alpha \matB_{n-1}\boldv_{n-1},  \label{vrecD}
\end{equation}
where $\matD_n=\C_n+\M_{n+1}\matB_n$, allowing all $\boldv_n$ to be calculated by simple recurrence.

The $s\to\infty$ asymptotic expressions for both the slow and Alfv\'en waves, both upgoing and downgoing, are now fully specified:
\begin{equation}
\U \sim  s^{1/2+2i\kappa\cos\phi\tan\theta}e^{\pm2i s\sec\theta}\sum_{n=0}^\infty \boldv_n s^{-n}  \mbox{ as $s\to\infty$, with the $+$ sign corresponding to downgoing waves}.
\end{equation}

For convenience, we present
\begin{multline}
\boldv_1 =
\pm\biggl(-\frac{1}{2} i p \cos \theta ,\ -\frac{1}{2} i q \cos \theta ,\ 0,\\
\qquad\frac{i \left(p \cos \theta 
   \left(\gamma ^2 \left(8 \nu ^2+3\right)+\frac{16 (\gamma -1) \kappa ^2}{\nu ^2}\right)+8 \gamma 
   \left(-\gamma  \nu ^2 p \sec \theta +\sin \theta  \left(-\gamma  \nu ^2 \tan \theta \, (p \cos 2
   \phi +q \sin 2 \phi )-2 i \kappa  (\gamma  p \cos \phi -q \sin \phi )\right)\right)\right)}{16
   \gamma ^2},\\
  \qquad \frac{\kappa  \sin \theta  (p \sin \phi +\gamma  q \cos \phi )}{\gamma }-i \nu ^2 \sin
   \theta  \tan \theta  \sin \phi \, (p \cos \phi +q \sin \phi )+\frac{3}{16} i q \cos \theta
   ,\ \kappa  p \cos \theta \biggr)^T\\
\end{multline}
and the  $\bxi$ part of $\boldv_2$
\begin{multline}
\boldv_2 = \left(\frac{p \cos ^2\theta  \left(\gamma ^2 \nu ^2 \left(8 \nu ^2-1\right)+16 (\gamma -1) \kappa
   ^2\right)-8 \gamma  \nu ^2 \left(\gamma  \nu ^2 p+\gamma  \nu ^2 \sin ^2\theta  (p \cos 2 \phi +q
   \sin 2 \phi )-i \kappa  q \sin 2 \theta  \sin \phi \right)}{32 \gamma ^2 \nu ^2},\right. \\
 \left.  \qquad  -\frac{1}{32} q
   \cos ^2\theta -\frac{\sin \phi  \left(2 \gamma  \nu ^2 \sin ^2\theta \, (p \cos \phi +q \sin \phi
   )+i \kappa  p \sin 2 \theta \right)}{4 \gamma },
\  -\frac{1}{2} i \kappa  p \cos ^2\theta,\, \ldots, \, \ldots, \,\ldots\right)^T.\\
\end{multline}

The coefficients become prohibitively complex as $n$ increases, and are best calculated numerically. Typically, 10 or 20 terms in these series are used in our numerical calculations at $s\approx10$, subject to the optimal truncation rule of asymptotic series \citep{BenOrs78aa}. We often truncate before this stage, when the terms are already adequate ($\|\boldv_n\|/s^n <10^{-8}\|\boldv_0\|$).

\section{Energy Flux}  \label{sec:Fz}
The vertical component of the wave-energy flux can be written as a Hermitian form, $F_z=F_0 \,\U^\dag  \Z\, \U$, where 
 {\footnotesize
\begin{multline}  \label{Z}
\Z=\\
 \begin{pmatrix}
    0 & -\kappa  \nu  \sin \theta  \cos \theta  \sin \phi  & \kappa  \nu  \sin ^2\theta  \sin ^2\phi
   +\frac{\kappa  s^2}{\nu } & \frac{1}{2} i \nu  \cos ^2\theta  & 0 & -\frac{1}{2} i \nu  \sin \theta
    \cos \theta  \cos \phi  \\
 -\kappa  \nu  \sin \theta  \cos \theta  \sin \phi  & \kappa  \nu  \sin 2 \theta \cos \phi  &
   -\kappa  \nu  \sin ^2\theta  \sin \phi  \cos \phi  & 0 & \frac{1}{2} i \nu  \cos ^2\theta  &
   -\frac{1}{2} i \nu  \sin \theta  \cos \theta  \sin \phi  \\
 \kappa  \nu  \sin ^2\theta  \sin ^2\phi +\frac{\kappa  s^2}{\nu } & -\kappa  \nu  \sin ^2\theta 
   \sin \phi  \cos \phi  & 0 & -\frac{1}{2} i \nu  \sin \theta  \cos \theta  \cos \phi  &
   -\frac{1}{2} i \nu  \sin \theta  \cos \theta  \sin \phi  & \frac{i \left(\nu ^2 \sin ^2\theta
   +s^2\right)}{2 \nu } \\
 -\frac{1}{2} i \nu  \cos ^2\theta  & 0 & \frac{1}{2} i \nu  \sin \theta  \cos \theta  \cos \phi 
   & 0 & 0 & 0 \\
 0 & -\frac{1}{2} i \nu  \cos ^2\theta  & \frac{1}{2} i \nu  \sin \theta  \cos \theta  \sin \phi 
   & 0 & 0 & 0 \\
 \frac{1}{2} i \nu  \sin \theta  \cos \theta  \cos \phi  & \frac{1}{2} i \nu  \sin \theta  \cos
   \theta  \sin \phi  & -\frac{i \left(\nu ^2 \sin ^2\theta +s^2\right)}{2 \nu } & 0 & 0 & 0 \\
      \end{pmatrix},
\end{multline}}
which is Hermitian, $\Z^\dag =\Z$. The superscript $\dag$ denotes the conjugate transpose. This total flux $F_z$ is independent of $z$, as may be verified by noting that $s F_z'=F_0\,\U^\dag (s\Z'+\Z \A+\A^\dag \Z)\,\U$, and checking that $s\Z'+\Z \A+\A^\dag \Z=0$. In practice, $F_0$ is chosen to set $F_z=1$ for the injected wave.


\bsp	
\label{lastpage}
\end{document}